\documentclass[prd, showpacs, superscriptaddress,eqsecnum, twocolumn,nofootinbib,preprintnumbers]{revtex4}
\usepackage[utf8]{inputenc}
\usepackage{bm}
\usepackage{latexsym,amssymb,amsmath,amsfonts}
\usepackage{graphicx}
\usepackage[usenames]{color}
\usepackage[breaklinks, backref, colorlinks=true]{hyperref}
\usepackage{epstopdf}

\def\d{\textrm{d}}
\def\eff{\textrm{eff}}
\def\LMW{L_\text{MW}}
\def\DelOm{\Delta\Omega}
\def\Virgo{\text{V}}
\def\iso{\text{iso}}
\def\Om{\mathbf{\hat{\Omega}}}
\def\ft{\widetilde}
\def\erf{\mathrm{erf}}
\def\P{\mathcal{P}}
\def\T{\delta T}
\def\true{\text{true}}
\def\obs{\text{obs}}
\def\SNR{\text{SNR}}

\graphicspath{ {fig/} }

\begin{document}

\preprint{IUCAA-02/2014, LIGO-P1300223}


\title{Astrophysical motivation for directed searches for a stochastic gravitational wave background}

\author{Nairwita Mazumder}
\email[email: ]{nairwita@iisertvm.ac.in}
\affiliation{School of Physics, IISER-Thiruvananthapuram, CET Campus, Trivandrum 695016, India}

\author{Sanjit Mitra}
\email[email: ]{sanjit@iucaa.ernet.in}
\affiliation{IUCAA, P. O. Bag 4, Ganeshkhind, Pune 411007, India}

\author{Sanjeev Dhurandhar}
\email[email: ]{sanjeev@iucaa.ernet.in}
\affiliation{IUCAA, P. O. Bag 4, Ganeshkhind, Pune 411007, India}

\begin{abstract}
The nearby universe is expected to create an anisotropic stochastic gravitational wave background (SGWB). Different algorithms have been developed and implemented to search for isotropic and anisotropic SGWB. The aim of this paper is to quantify the advantage of an optimal anisotropic search, specifically comparing a point source with an isotropic background. Clusters of galaxies appear as point sources to a network of ground based laser interferometric detectors. The optimal search strategy for these sources is a ``directed radiometer search''. We show that the flux of SGWB created by the millisecond pulsars in the Virgo cluster produces a significantly stronger signal than the nearly isotropic background of unresolved sources of the same kind. We compute their strain power spectra for different cosmologies and distribution of population over redshifts. We conclude that a localised source, like the Virgo cluster, can be resolved from the isotropic background with very high significance using the directed search algorithm. For backgrounds dominated by nearby sources, up to redshift of about 3, we show that the directed search for a localised source can have signal to noise ratio more than that for the all sky integrated isotropic search.
\end{abstract}

\maketitle

\section{Introduction}

Einstein's theory of general relativity predicts gravitational waves (GWs)~\cite{gw,MTW}. The existence of GWs was first confirmed by the observation of decay in the orbital period of the Hulse-Taylor binary pulsar system (PSR B1913+16)~\cite{HulseTaylor,WeisbergTaylor}.  Recently a very strong claim of detection of imprints of primordial GWs on Cosmic Microwave Background polarisation was made~\cite{BICEP2-01}. While ``direct'' detection of GWs has not been possible yet, the first generation ground based Laser Interferometric detectors, LIGO~\cite{LIGO,LIGORev}, Virgo~\cite{Virgo,VirgoRev}, GEO600~\cite{GEO600,GEORev} and TAMA300~\cite{TAMA300,TAMARev}, have demonstrated the capability to measure strain signal of the order $10^{-23}$. However, the second generation detectors, which are currently being installed~\cite{AdvLIGO,AdvVirgo,KAGRA}, will have ten times greater sensitivity and will cover a broader frequency spectrum. This will enable the detectors to  observe at least three orders of magnitude more volume, thereby enhancing the chances of detection  from {\em few percent level to close to unity} in the next few years.

Sources of GWs can be broadly classified in three categories based on their duration and phase coherence:
\begin{enumerate}
\item Burst sources: Short duration sources with modelled (e.g., compact binary coalescence) or unmodelled (e.g., supernovae) phase evolution.
\item Continuous sources: long duration sources {\em with} phase coherence (e.g.,  spinning neutron stars).
\item Stochastic Background: created by a collection of unresolved and independent sources {\em without} phase coherence (e.g., coalescing binaries, millisecond pulsars in a galaxy cluster).
\end{enumerate}
Coalescing compact binary stars are the most promising sources of GWs, as the waveform from these sources can be modelled to a very high degree of accuracy, which makes it possible to apply the techniques of matched filtering for digging out signal from noisy data from the detectors. However, GW astronomy promises a much broader spectrum of sources and many, if not most, of these waveforms will not be known a priori. The stochastic gravitational wave background (SGWB), by definition, is one such type of sources.

Different SGWBs are expected to be created from early universe phenomena~\cite{Grishchuk00,BinCap12} as well as from collection of astrophysical sources in the older low redshift universe. Here we only consider the astrophysical background, which can be generated by an incoherent superposition of short and long duration sources. These sources can be stochastic either in the time domain or in the frequency domain. In the time domain it can appear as pop-corn noise, a large set of events non-uniformly distributed over a certain interval of time, e.g., a population of supernovae, rotating neutron stars (including pulsars, magnetars and gravitars),  binary super-massive black holes  in a galaxy cluster~\cite{TanFre03,Marassi11,CowTan06,Zhu12, Rosado11, RaviWyi12, Mingarelli13,Palomba05}. In the frequency domain the background can be created by a ``forest of emission lines'', narrow or broad, whose exact frequencies are not known, but the distribution can be modelled. For instance, a population of millisecond pulsars in a cluster of galaxies can create such a background~\cite{TanFre06, hotspot}. 

Detection of these astrophysical SGWBs can provide collective information about the constituent sources which are not accessible by conventional electro-magnetic (EM) observations. In particular, average physical properties of such sources, such as the mass asymmetry of neutron stars, the equation of state, the population distribution etc. can be probed via SGWB observations.

The best strategy to search for stochastic signals is by cross-correlating signals from different detectors. Over the last three decades algorithms have been developed and implemented to search for  isotropic and different kinds of anisotropic SGWBs~\cite{Michelson87, christ92, flan93, allen01, allen97, LIGOT04014000Z, kudohI, kudohII, taruyaIII, ballmer06, mitra08, mitra09a, Mitra10b, LSC05a, LSC06b, sgwbS4dir, sgwbS5iso,sgwbS5dir}.  Further, the expected signal-to-noise ratios (SNRs) from optimized searches for anisotropic SGWBs were computed. However, the relative strengths of the expected isotropic and anisotropic backgrounds, based on the current knowledge of astronomy, and the corresponding relative SNRs, have not been studied in literature. Such a study would in turn provide a firm justification for performing (or not performing) dedicated searches for different isotropic and anisotropic backgrounds.

The picture of the universe that we get from EM astronomy shows that the nearby universe is highly anisotropic, while at large scales it is fairly isotropic. Hence, the true background will contain an isotropic as well as an anisotropic component. Our first task will be to compare their relative strengths.

In the analysis of data from GW detectors, the searches for isotropic SGWBs are motivated from the fact that the primordial SGWB is statistically isotropic (though different models for power spectral densities are still needed to search for different astrophysical sources which create the background). An anisotropic SGWB search will be justified in a situation provided {\em either} of the following criterion is satisfied:
\begin{enumerate}
\item the anisotropic search can confidently probe the anisotropy, i.e., the search should be able to clearly distinguish the anisotropic component from the isotropic part;
\item the (optimal) anisotropic search has a significantly large SNR as compared to the (suboptimal) isotropic search.
\end{enumerate}
Because the nearby universe dominates the background, one could na\"ively anticipate that the above criteria are always satisfied.  However, as we show in this paper, this is not necessarily the case. Nearby sources may appear strong because of their proximity, but the distant sources are large in number. In general, the comparison depends on the population distribution of the sources at different frequencies and redshifts and the expansion history of the universe. Here we quantify the relative strengths of these backgrounds and discuss in which astrophysical situations the differences will be significant.

The specific case we consider in this paper is simple. We compare the relative strengths and detectability of a localised point source with those for an isotropic background created by similar sources with correct (optimal) and interchanged (suboptimal) filters. In our example the anisotropic part is created by a large number of Milli-Second Pulsars (MSPs) in the Virgo cluster~\cite{hotspot}. The Virgo cluster~\cite{VirgoCluster} is a localised source which can be assumed to be a point source of SGWB. This assumption is justified because the angular width of the Virgo cluster (few degrees) is comparable to the angular resolution of the network of present ground based Laser Interferometric GW detectors~\cite{mitra08}. The isotropic part is created by all such MSPs in the rest of the universe whose distribution is nearly isotropic. We first show that the total GW flux received from the Virgo cluster exceeds the flux {\em from the same solid angle as that of the Virgo cluster} from the integrated distant isotropic universe. This is similar to solving the ``Olbers' paradox''~\cite{invernoGR} problem in cosmology. However, this calculation would provide the comparison between the SNRs observed in different searches, if the searches had uniform frequency response and infinite bandwidth, which is of course impossible. To perform a realistic comparison, we evaluate the redshift integrated spectra of the background for different cosmological evolution and (simple) population distribution models. Combining these spectra with the frequency response of the search one can obtain the final observed SNRs. We compare the SNRs that one would observe for directed and isotropic searches for both the localised and isotropic components of the background. The computed numbers can then be used to draw a final conclusion regarding the effectiveness of different searches in varied cosmological conditions.

The paper is organised as follows. In section~\ref{sec:olbers} we provide a brief review of the  Olbers' paradox and its more general form in terms of frequency spectrum. We compare the fluxes and the frequency spectra for a localised source and the isotropic background in section~\ref{sec:background}. In section~\ref{sec:SNR} we compute the numerical results for the observed SNR for different combinations of sources and models. We discuss the results and future directions in section~\ref{sec:conclusion}.

\section{Olbers' Paradox}
\label{sec:olbers}

A common observation is that the night sky is mostly dark. The stars and galaxies stand out in the night sky as the background is almost negligible. This observation can be reconciled with the physics by the apparently straightforward argument that the observed sky should be dominated by the nearby universe. However this argument is flawed. Actually, the number of sources in the universe per unit solid angle increases exactly in the same way as their flux received at the earth reduces, thereby compensating the effect of distance. To pose the problem mathematically, let us consider a solid angle $\Delta\Omega$. If the universe is homogeneous, the number of sources between distance $r$ and $r + \d r$ that are contributing to the flux in this solid angle is given by $n  r^2 \Delta\Omega \d r$, where $n$ is the average number density of sources in the universe. However the flux from these sources reduces as $1/r^2$.  Hence, the ``effective'' number of contributing sources, which would produce the same amount of flux at the point of observation when placed unit distance away, is proportional to $n \Delta\Omega \d r$ and does not depend explicitly on $r$. That is, every distant part of the universe would contribute equally to this solid angle $\Delta\Omega$, which would make the night sky almost uniformly bright in an isotropic universe and, infinitely bright in an infinite universe. The seeming incongruity of this prediction with our observations of the night sky is Olbers' paradox. In case of GWs, Olbers' paradox would imply that the anisotropic SGWB would be insignificantly small as compared to the isotropic part, as the latter is created by the distant universe which is much deeper than the local universe.

The solution to Olbers' paradox becomes obvious when one includes the expansion of the universe. According to general relativity, GWs propagate in the same way as EM waves along null geodesics~\cite{MTW}. In an expanding universe described by the homogeneous and isotropic Friedmann-Lema\^itre-Robertson-Walker (FLRW) metric~\cite{narlikarCosmo}, along the null geodesics leading to an observer at the origin of the coordinate system, the following condition is satisfied 
\begin{equation}
c^2 \, \d t^2 \ - \ a^2(t) \, \d r^2 \ = \ 0 \, .
\end{equation}
where $t$ is the time-like coordinate, $r$ is the space-like radial coordinate and $a(t)$ is the so called {\em scale factor}.
Thus, if two pulses are emitted by a source at radial coordinate $r$ at time $t_1$ and $t_1 + \delta t_1$, which reach the observer at the current epoch $t_0$ and $t_0 + \delta t_0$ respectively, one can write
\begin{equation}
r \ = \ c \int_{t_1}^{t_0} \frac{\d t}{a(t)} \ = \ c \int_{t_1 + \delta t_1}^{t_0 + \delta t_0} \frac{\d t}{a(t)} \, ,
\end{equation}
The above relation yields $\delta t_0 = a(t_0)/a(t_1) \, \delta t_1$. This implies that the rate at which the pulses are observed and energy received is {\em lower} by a factor of $a(t_0)/a(t_1) =: 1 + z$, where $z$ is the usual cosmological redshift. Moreover, due to the expansion of the universe the wavelengths are stretched and hence energy of each pulse is again reduced by the same factor of $a(t_0)/a(t_1) = 1 + z$. Combination of these two factors implies that the flux received from a source of luminosity $L$ (total energy released per unit time as measured by an observer very close to the source), at a redshift $z$ is~\cite{narlikarCosmo}
\begin{equation}
F \ = \ \frac{L}{4 \pi r^2 a^2(t_0) \, (1+z)^2} \, . \label{eq:flux}
\end{equation}
Thus expansion of the universe reduces the flux and, hence the effective number of sources, by a factor of $(1+z)^2$. Then the effective number of sources between comoving distance $r$ and $r + \d r$ becomes $n \, \Delta\Omega \, \d r / (1 + z)^2$, where $n$ is the {\em comoving} source number density\footnote{If the universe was expanding but the structures did not evolve, the comoving number density would be a constant.} at redshift $z$. Since the redshift $z$ is a monotonically increasing function of coordinate distance $r$,  the distant universe appears more and more dim. In practice, the total number of galaxies in the universe also reduces with redshift, which plays a role in reducing the above number even further. However, this has a weaker effect, because at high redshifts ($z \gtrsim 2$) the effective number density gets a very low weightage due to the $1/(1+z)^2$ factor.

The solution to Olbers' paradox for EM astronomy already implies that the same will hold for SGWB, as there is no difference between these two in this context. GW flux reduces in the same way as for EM waves~\cite{MTW, ThorneBook}. Hence, the extragalactic astrophysical GW background is created mostly by low-to-intermediate redshift sources.

The above resolution for total flux will however suffice if the detection scheme had flat frequency response over the whole infinite frequency range. This is, of course, never true in practice. So what is more relevant for studying the detectability of different backgrounds is a more general quantity than the flux, namely, the observed frequency spectra $S(f)$. The total flux $F$ is the integrated value of $S(f)$ over the entire frequency range.
If a source with luminosity $L$ at a redshift $z$ had an intensity distribution $J(f)$, such that $\int  J(f) \, \d f = 1$, one can show by extending the derivation of Eq.~(\ref{eq:flux}) and taking into account the redshift of frequency interval $\d f$ that the observed spectrum is~\cite{narlikarCosmo,weinbergCosmo},
\begin{equation}\label{eq:sf1}
S(f) \ = \ \frac{L \, J(f(1+z))}{4\pi r^2 \, a^2(t_0) \, (1+z)} \, .
\end{equation}
The above formula will be the starting point for the main strain power spectrum density (PSD) $H(f)$ calculation done in the next section. Note that the frequency integral of the above expression, the total flux,
\begin{equation}
F \ = \ \int S(f) \, \d f\ \ = \ \frac{L \int \d f  J(f(1+z))}{4\pi r^2 \, a^2(t_0) \, (1+z)} \, ,
\end{equation}
identically matches the expression in Eq.~(\ref{eq:flux}).

\section{Localised vs. the Isotropic background}
\label{sec:background}

In this section we compare the GW flux and power spectra generated by the Virgo cluster with those from the rest of the universe for the same kind of source. While the analytical framework we have developed here is valid for any kind of source, for the numerical evaluation we have used the milli-second pulsars (MSPs). This is because the strain PSD of the SGWB created by MSPs in the Virgo cluster is available as a ready result~\cite{hotspot}. Also note that the overall amplitude of the strain due to optimistic or pessimistic assumptions (e.g., about the mass-asymmetry of the neutron stars) {\em do not matter in this work}, as we are only interested in the relative strengths of the localised and isotropic background created by the same kind of sources. So the strength of individual sources cancel out in the calculation.

A typical Milky Way like galaxy is expected to have at least $40,000$ MSPs~\cite{Lorimer-rev}. Each of the MSPs is expected to emit narrow band GW signal~\cite{mspgw}, which in total would appear as a forest of emission lines on the frequency axis. In a galaxy cluster like the Virgo with an estimated $\sim 10^{8}$ MSPs, the forest is so dense that it appears as a continuum. Using a population distribution model for the MSPs, the SGWB from the Virgo cluster was computed in \citet{hotspot}. Here we essentially integrate their results over different redshifts to get the PSDs, which in turn, gives the SNR for different searches.

It is important to emphasise that the GW background considered here is created by {\em spinning neutron stars} with mass distribution not symmetric about the spin axis and with period of few milliseconds, whether or not they emit EM ``pulses''. The population models based on EM observations provide an estimate for MSPs, though there can be many more spinning neutron stars without any EM emission, e.g., gravitars~\cite{Palomba05}. In this paper, since we are interested in relative fluxes and SNRs, the numerical results only depend on the distribution of the sources, {\em not on their total number}. We assume the estimated distribution of MSPs is applicable to the whole set of spinning neutron stars. Had the total number of such sources been different, {\em all} the power spectra and the SNRs would have a different, but common, scaling factor, which does not affect the conclusions.

\subsection{Comparison of total Flux}

In the context of Olbers' paradox, the first quantities we compare are the total flux received from the Virgo cluster with that received on the average from the same solid angle as the cluster in an otherwise statistically isotropic universe. Instead of quoting numbers for flux, we quote the effective number of Milky Way Equivalent Galaxies~(MWEG), $N_\eff$, which would produce the same amount of flux unit distance away. We find this quantity to be more intuitive than flux. $N_\eff$ is essentially flux ($F$) in different units, related by the formula
\begin{equation}
F \ = \ N_\eff \, \frac{\LMW}{4 \pi} \, ,
\end{equation}
where $\LMW \sim 1.5 \times 10^{10} L_{\odot}$ is the blue luminosity of a MWEG and  the solar luminosity $L_{\odot} \approx 4 \times 10^{33} \text{erg}  \, s^{-1}$. Thus the ratio of fluxes is the same as the ratio of $N_\eff$, which is our primary interest in this paper. Note that, $N_\eff$ has the dimensions of inverse distance square. This is because if the unit of distance increases, say from Mpc to Gpc (= $10^3$ Mpc), $N_{\eff}$ must increase by a factor of $10^6$ to maintain the same flux.

Since the Virgo cluster is $d_\Virgo = 16.5$ Mpc luminosity distance away (which is also the same as the comoving distance when $a(t_0)$ is set to unity for a very low redshift source like the Virgo cluster) and it has about $N_\Virgo = 1500$ MWEG~\cite{VirgoCluster}, $N_\eff$ for Virgo cluster is
\begin{equation}
N_\eff^\Virgo \ = \ N_\Virgo / {d_\Virgo}^{2} \ \approx \ 5.5 \, \text{Mpc}^{-2}. \label{eq:neffvirgo}
\end{equation}

We now compute the average $N_\eff$ in a statistically isotropic universe from the solid angle subtended by the Virgo cluster $\DelOm_\Virgo = 0.012$ steradian ($\sim 50$ sq. degree, equivalent to a circular region of radius $\sim 8$ degree). For this we would need to integrate over the radial distance coordinate $r$ and account for cosmological redshift of the emitted waves.

If the {\em comoving} number density of MWEG in the universe in the Virgo solid angle $\DelOm_\Virgo$ is a function of redshift, given as $n(z)$, the total flux received by the observer can be obtained by integrating Eq.~(\ref{eq:flux}) over spherical shells as
\begin{equation} \label{eq:isoFlux}
F_\iso \ = \ \frac{\LMW}{4 \pi} \, \int_0^\infty \d r \, \frac{\DelOm_\Virgo r^2 \, n(z)}{r^2 a^2(t_0) \, (1+z)^2} \, .
\end{equation}
After simplifying, this can be written in terms of $N_\eff$ as
\begin{equation}
\label{eq:Neffr}
N_\eff^\iso \ = \ \DelOm_\Virgo \, \int_0^\infty \d r \, \frac{n(z)}{a^2(t_0) \, (1+z)^2} \, .
\end{equation}
In order to compute the above integral analytically or numerically, we need to relate the coordinate distance $r$ with redshift $z$. We use the standard relation~\cite{peacockCosmo},
\begin{equation}
\label{eq:drdz}
\d r \ = \ (c/H_0) \, \d z / E(z) \, ,
\end{equation}
where, $H_0$ is the Hubble constant and, in a flat universe,
\begin{equation}
\label{eq:defE}
E(z) := \sqrt{\Omega_\Lambda + \Omega_m(1+z)^3} \, .
\end{equation}

In this paper, we assume that co-moving number density of MWEG is not evolving with time, $n(z) = n_0$, where $n_0$ is the current galaxy number density of the universe in the units of MWEG/volume. Instead of choosing a ``top hat'' model for $n(z)$, we could consider a detailed number density versus redshift relation, which could provide more accurate source PSD for that model. However, our primary interest is not to get accurate numbers for a particular type of model, our aim is to show that there can be anisotropic SGWB created by sources in the nearby universe, with certain density distribution over redshift, for which the directed search and isotropic search can provide non-overlapping information.  So a simplistic model, like the one used here, is preferred in this context. Moreover, we are considering localised sources created by distributions of MSPs which are more likely to be present in the nearby universe where the co-moving matter density can at most evolve slowly until $1/(1+z)^2$ becomes too small. So a nearly constant profile with a redshift cut-off $z = z_\text{max}$ does indeed provide a reasonable model. Nevertheless, in the rest of this paper we derive general formulae valid for any arbitrary $n(z)$, only to get the final results we use
\begin{equation}
n(z) \ = \ \left\{
\begin{array}{cl}
n_0 & \text{if} \ z \le z_\text{max} \\
0 & \text{otherwise} \, .
\end{array}
\right.
\label{eq:nofz}
\end{equation}

Given the redshift cutoff discussed above and setting the scale factor at current epoch $a(t_0) = 1$ without any loss of generality, the form of $N_\eff^\iso$ becomes,
\begin{equation}
\label{eq:Neff}
N_\eff^\iso \ = \ n_0 \, \DelOm_\Virgo \, \frac{c}{H_0} \, \int_0^{z_\text{max}} \frac{\d z}{(1+z)^2 \, E(z)} \, .
\end{equation}
We now evaluate the above expression in the specific cases. Here we consider only flat universe ($\Omega_\Lambda + \Omega_m = 1$) and compute the results for two different cosmologies:
\begin{enumerate}
\item Universe without dark energy

A flat matter dominated universe without dark energy ($\Omega_\Lambda = 0, \Omega_m = 1$). This case allows us to get the results analytically. These results may not be very realistic, but they provide useful insights through their analytical forms.

In this case eqn.~(\ref{eq:Neff}) becomes
\begin{eqnarray}
N_\eff^\iso &=& \DelOm_\Virgo \, \frac{c}{H_0} \, \int_0^{z_\text{max}} \d z \, \frac{n_0}{(1+z)^{7/2}} \\
&=&  n_0 \, \DelOm_\Virgo \frac{2}{5} \, \frac{c}{H_0} \left[ 1 \, - \, \frac{1}{(1+z_\text{max})^{5/2}} \right].
\end{eqnarray}
For $z_\text{max} \rightarrow \infty$, assuming $H_{0} \sim 72 \text{km/sec/Mpc}$  and $n_0 \sim 0.01~h^3~\text{Mpc}^{-3} \sim 0.004~\text{Mpc}^{-3}$~\cite{longairGalaxy,peeblesCosmo}, $N_\eff^\iso$ becomes $\sim 0.082 ~\text{Mpc}^{-2}$.
Hence, by comparing with eqn.~(\ref{eq:neffvirgo}), one can conclude that the Virgo cluster is $5.5 / 0.08 \approx 69$ times brighter than the statistically isotropic background in this cosmology.

\item $\Lambda$CDM cosmology

The ``standard'' Lambda Cold Dark Matter ($\Lambda$CDM) universe with parameters taken from observational cosmology ($\Omega_\Lambda \approx 0.73, \Omega_m \approx 0.27$).

In this case the integral in eqn.~(\ref{eq:Neff}) can be evaluated numerically. Here, for $z_\text{max} \rightarrow \infty$,  the contrast factor turns out to be $\sim 50$.
\end{enumerate}
In general, the above contrast ratios are functions of redshift cut-off $z_\text{max}$. Table~\ref{tab:contrast} lists them for few relevant values of $z_\text{max}$. Note that the results for $z_\text{max} = 10$ and $z_\text{max} = \infty$ are provided mainly for academic interest, as MSPs are very unlikely to exist beyond $z_\text{max} \sim 3$ and the top-hat model, $n(z) = n_0$, cannot hold for such high values of $z_\text{max}$. On the other hand, the presence of the $(1+z)^{-2}$ factor in Eq.~(\ref{eq:isoFlux})  ensures that even in those extreme $z_\text{max}$ cases, though the PSDs shown in Fig.~\ref{fig:H} differ significantly, the total flux is minimally affected as seen in Table~\ref{tab:contrast}. However, $z_\text{max}$ does affect the SNRs for isotropic search, as computed in the subsequent sections.
\begin{table*}[t]
\begin{tabular}{|l|r|r|r|r|r|r|r|r|r|r|r|}\hline\hline
& \multicolumn{5}{|c|}{No dark energy} & \multicolumn{5}{|c|}{$\Lambda$CDM cosmology} \\\cline{2-11}
$z_\text{max}$ & $1$ & $2$ & $3$ & $10$ & $\infty$ & $1$ & $2$ & $3$ & $10$ & $\infty$\\\hline
$N^\iso_\eff ~(\text{Mpc}^{-2})$ & 0.0659 & 0.0749 & 0.0775 & 0.0798 & 0.0800 & 0.0834 & 0.0991 & 0.1040 & 0.1084 & 0.1088 \\\hline
$N^\Virgo_\eff / N^\iso_\eff$ & 83.5 & 73.4 & 71.0 & 68.9 & 68.7 & 65.9 & 55.5 & 52.9 & 50.7 & 50.5 \\\hline\hline
\end{tabular}
\caption{The first row of this table lists the effective number of MWEG sources $N^\iso_\eff$ placed $1$Mpc away from the detector, such that the total flux is equivalent to that from all the sources present in the same solid angle as the Virgo cluster. The second row shows the ratio of the effective number of sources in the Virgo cluster, $N^\Virgo_\eff$, to $N^\iso_\eff$, which essentially quantifies how ``bright'' the Virgo cluster would appear in the isotropic background, when observed with an instrument with a flat frequency response and infinite bandwidth. Note that $N^\Virgo_\eff = 5.5 ~\text{Mpc}^{-2}$. \label{tab:contrast}}
\end{table*}

Although the above exercise explains Olbers' paradox specifically for the Virgo cluster, it is true only in principle, for a detector which captures the whole spectrum of the (redshifted) sources with uniform response. In practice, the observed contrast could be more or less than this value depending on the spectrum of the source and the frequency response of the detector, which will be the main consideration in the rest of the paper.

Note that these results and the above discussion do not involve anything special about GW, they are equally valid for EM observations. The only difference arises in the definition of the response function. In EM astronomy the detectors generally count photons, while the GW detectors measure strain. However, the strain response function of a GW detector can be converted to flux response function, which will have different frequency response, but  the forms of the expressions remain the same. Thus, in some sense, the analytical treatment developed below for GW, can be extended in a straightforward manner to include EM detectors.

\subsection{Comparison of Power Spectral Densities (PSDs)}

Redshift not only reduces intensity, but it shifts power lower to frequencies, which essentially means that the further the sources are the lower the frequency where their power is concentrated. In this subsection we explicitly derive an expression for the strain PSD of the integrated SGWB generated at different cosmological distances and compare it with the Virgo-only PSD derived in \citet{hotspot}.

Let $J(f)$ be the intensity distribution of the SGWB created by the sources (e.g., MSPs) in a MWEG. As given in ({\ref{eq:sf1}}), for the source at redshift $z$, the GW \textit{flux density} near the earth is,
\begin{equation}
\label{eq:sf}
S(f) \ = \ \frac{\LMW \, J(f(1+z))}{4\pi r^2 \, a^2(t_0) \, (1+z)} .
\end{equation}

SGWB strain signal $h_{ij}$ at a given point in space $\mathbf{x}$ and time $t$ can be expanded in Fourier modes as
\begin{align}
\label{eq:plwave}
h_{ij}(t,\mathbf{x}) & \ = \\
\sum_{A=+,\times} & \int_{-\infty}^\infty \d f \int_{S^2} \d\Om \, \ft{h}_A(f,\Om) \, e^A_{ij}(\Om) \, e^{2\pi i f (t - \Om\cdot\mathbf{x}/c)} \, , \nonumber
\end{align}
where $\Om$ is the propagation direction of the wave, $f$ is the frequency and $e^A_{ij}(\Om)$ is the symmetric-trace-free (STF) basis tensors. The Fourier components, $\ft{h}_{+,\times}(f,\Om)$, have no correlations, for an SGWB us generated by a set of incoherent events, like the ones considered in this paper. This can be expressed as
\begin{align}
\langle \ft{h}_A(f,\Om) \, \ft{h}_{A'}(f',\Om')\rangle \ &= \nonumber \\
\delta_{AA'} \, \delta(f-f') &\, \delta^2(\Om - \Om') \, H(f) \, {\cal P}(\Om) \, ,
\label{eq:uncorrh}
\end{align}
where $H(f)$ represents the shape of the frequency power spectrum of the background and ${\cal P}(\Om)$ is the direction dependent amplitude of the power spectrum. Note that, in general the $f$ and the $\Om$ components need not be separable. That is, the shape of the spectrum can be different in different directions. So, in general one should use a function of the form $\bar{\cal P}(\Om, f)$ instead of $H(f) \, {\cal P}(\Om)$. However, this is not a relevant issue in this paper, ${\cal P}(\Om)$ and $H(f)$ are indeed separable for the case we consider here. This is because we are considering either the Virgo cluster, localised in the direction $\Om_\Virgo$, for which the spectra of the other directions are zero, so without any loss of generality one can write $\bar{\cal P}(\Om, f) = \delta^2(\Om - \Om_\Virgo) H_\Virgo(f)$. For an isotropic background, by definition, the spectra is the same in every direction, $\bar{\cal P}(\Om, f) = H_\iso(f)$. Thus, in both the cases  $\bar{\cal P}(\Om, f)$ is separable by construction.

Also note that, $H(f)$ and ${\cal P}(\Om)$ can be normalised keeping the product fixed. So, without any loss of generality one can impose one extra condition that,
\begin{equation}
\int \d \Om \, {\cal P}(\Om) = 1 \,  ,
\end{equation}
which means that ${\cal P}(\Om) = \delta^2(\Om - \Om_\Virgo)$ for the Virgo cluster and ${\cal P}(\Om) = 1/4\pi$ for an isotropic background.

The total energy density of GW is given by~\cite{MTW} 
\begin{equation}
 \rho_\text{GW} \ = \ \frac{c^2}{32 \pi G} \, \langle \dot{h}(\textbf{x},t) \, \dot{h}(\textbf{x},t) \rangle \, ,
\end{equation}
where an overdot represents derivative with respect to time in the observer's frame.
We are interested in finding the GW flux density from a localised source, specifically the Virgo cluster. In the above equation, substituting ${\cal P}(\Om) = \delta(\Om - \Om_\Virgo) $, eqn.~(\ref{eq:plwave}) and eqn.~(\ref{eq:uncorrh}), and comparing with the relation
\begin{equation}
\rho_\text{GW} \ = \ \frac{1}{c} \int_0^\infty \d f \, S(f) \, ,
\end{equation}
one can arrive at the formula,
\begin{equation}
S(f) \ = \ \frac{c^3}{32 \pi G} \, f^2 \, H(f)\, .
\end{equation}
Finally, combining this with eqn.~({\ref{eq:sf}}) one can finally write the expression for $H(f)$ for a $N_\text{gal}$ localised MWEG in a given direction as
\begin{equation}
\label{eq:oneSourceH}
H(f) \ = \ N_\text{gal} \, \frac{8 G \,  L_\text{MW} \, J(f(1+z))}{c^3 \, f^2 \, r^2 \, a^2(t_0) \, (1+z)} \, .
\end{equation}
For the Virgo cluster, $N_\text{gal} \ := \ N_\Virgo \ \approx \ 1500$ and $z \approx 0$. Hence,
\begin{equation}
\label{eq:hVirgo}
H_\Virgo(f) \ = \ N_\Virgo \, \frac{8 G \,  L_\text{MW} \, J(f)}{c^3 \, f^2 \, d_\Virgo^2} \, .
\end{equation}
However, the expression for {\em expected} $H_\Virgo(f)$ was already calculated in \citet{hotspot} for a given distribution of the pulsars $N(f)$ in a MWEG as,
\begin{equation}
\label{eq:hotSpotVirgo}
H_\Virgo(f)  \ = \ \tilde{h}_0^2 \, f^4 \, N(f) \, .
\end{equation}
The frequency independent constant $\tilde{h}_0$ is the typical strength of GW emitted by a $1.4 M_\odot$ neutron star:
\begin{equation}
\tilde{h}_0 \ \approx \ 7 \,  \langle {\alpha}^2 \rangle \, \times \, 10^{-34} \, \left(\frac{\epsilon}{10^{-5}}\right) \, \left(\frac{I}{1.1 
\times 10^{45} \text{g cm}^2}\right) \, .
\end{equation}
where $\epsilon$ is the ellipticity of the neutron star, $I$ is the moment of inertia, $\alpha \leq 1$ is the 
orientation factor and $\langle {\alpha}^2 \rangle$ represents the average with respect to the inclination
 angle and the polarisation angle. For uniformly distributed source over the angles $\langle {\alpha}^2 \rangle=2/5.$ We use this relation to alleviate the complications in fixing the proportionality constants as discussed below.

To compute the SGWB from the isotropic part, one has to integrate over all the sources in the same solid angle as what the Virgo cluster subtends at the earth. Since the number of MWEG $\d N_\text{gal}$ in solid angle $\DelOm_\Virgo$ in radial coordinate range $r$ to $r + \d r$ is $n(z) \DelOm_\Virgo r^2 \d r$, the net result can be obtained by integrating eqn.~(\ref{eq:oneSourceH}) as
\begin{equation}
\label{eq:hIso}
 H_\iso(f) \ = \ \DelOm_\Virgo \int_0^\infty \d r \, n(z) \, \frac{8 G \,  L_\text{MW} \, J(f(1+z))}{c^3 \, f^2 \, a^2(t_0) \, (1+z)} \, .
\end{equation}
Then invoking the relation between $\d r$ and $\d z$, eqn.~(\ref{eq:drdz}), one can arrive at the relation
\begin{equation}
H_\iso(f) \ = \ \DelOm_\Virgo \frac{c}{H_0} \int_0^\infty \d z \, \frac{8 G \,  L_\text{MW} \, J(f(1+z)) \, n(z) }{c^3 f^2 a^2(t_0) (1+z) \, E(z)} \, .
\end{equation}

Since we are primarily interested in comparing the relative strengths of the Virgo cluster and the isotropic background and we already have a formula for the Virgo cluster, we can hide the growing number of constants in the above equation, by dividing the above equation by eqn.(\ref{eq:hVirgo}). We can write
\begin{equation}
\frac{H_\iso(f)}{H_\Virgo(f)} \ = \ \frac{\DelOm_\Virgo \, d_\Virgo^2}{N_\Virgo} \frac{c}{H_0} \int_0^\infty \d z \, \frac{J(f(1+z)) \, n(z) }{a^2(t_0) J(f) \, (1+z) \, E(z)} \, .
\end{equation}
This further reduces complications involving the proportionality constants when we expand the intensity distribution $J(f)$ in terms of the source number distribution $N(f)$, in fact, a considerable amount of details were presented in \citet{hotspot} in fixing those constants.

We know that GW intensity is proportional to the square of triple derivative of quadrupole moment.  If there are $N(f) \d f$ sources in the GW frequency interval $f$ to $f + \d f$ in a Milky Way like galaxy, one should have $J(f) \propto N(f) \ f^6$, where the proportionality constant would be independent of frequency, and hence, redshift. Substituting this in the above equation and setting $a(t_0) = 1$, one gets
\begin{eqnarray}
\label{eq:ratIsoVirgo}
H_\iso(f)  &=& H_\Virgo(f) \ \left( \frac{\DelOm_\Virgo \, d_\Virgo^2}{N_\Virgo} \, \frac{c}{H_0} \right)  \, \times  \\
&& \quad \int_0^\infty \d z \, (1+z)^5 \frac{n(z)}{E(z)} \, \frac{N(f(1+z))}{N(f)} \, , \nonumber
\end{eqnarray}
where $H_\Virgo(f)$ is given by eqn.~(\ref{eq:hotSpotVirgo}).

To proceed further we would need to introduce the number distribution of the sources explicitly. Qualitatively this happens because the frequency distribution of sources at different redshifts is dependent on the actual distribution and the cosmology.

As mentioned earlier, in this paper we are explicitly evaluating quantities for MSPs in galaxy clusters and the rest of the universe. From the radio survey of our Galactic disk, the number of pulsars is estimated to be at least $40,000$. The population of these pulsars follow two different distributions, each distribution is Gaussian and has mean and standard deviation different from the others and they can be divided two regions separated by $50$Hz. The distributions in each region may be approximated as log-normal distributions as stated below~\cite{hotspot}. The probability that the log of pulsar rotation frequency is in the range $\log f_r$ and $\log f_r + \d \log f_r$ is given by
\begin{itemize}
\item for $f_r > 50$Hz
\begin{equation}\label{eq:p1}
p_1(\log f_r) \, \d\log f_r \ = \ \frac{1}{\sqrt{2 \pi} \sigma_1} \, e^{-\frac{{(\log f_r - \log \mu_1)}^2}{2 {\sigma}_1^2}} \d \log f_r \, ,
\end{equation}
\item for $f_r < 50$Hz
\begin{equation}\label{eq:p2}
p_2 (\log{f_r}) \, \d\log{f_r} \ = \ \frac{1}{\sqrt{2 \pi} \sigma_2} \, e^{-\frac{{(\log f_r - \log \mu_2)}^2}{2 {\sigma}_2^2}} \d \log f_r \, ,
\end{equation}
\end{itemize}
 where $\mu_1=219$Hz, $\sigma_1=0.238$, $\mu_2=1.71$Hz, $\sigma_2=0.420$ and $f_r=$Pulsar spin frequency $=$ half of gravitational wave frequency. Following \citet{hotspot}, here we consider a similar distribution of pulsars in the Virgo cluster and the rest of the universe. There are approximately $1500$  galaxies in the Virgo cluster. The total number of pulsars in our galaxy can be taken as $10^8$ for $f_r<50$ and $40,000$ for $f_r>50$. So the distribution of pulsars including MSPs in Virgo cluster becomes
\begin{equation}
\label{eq:nf}
N(f) \d f \ = \ [ N_\text{high} \, p_1(\log f_r) \ + \ N_\text{low} \, p_2(\log f_r)] \, \frac{\d f_r}{f_r \ln 10} \, ,
\end{equation}
with $N_\text{high} \sim 4 \times 10^7$ for $f_r>50$ and $N_\text{low} \sim  10^{11}$ for $f_r<50$. Note that the GW emission frequency $f$ from the MSPs is twice the rotation frequency $f_r$. For a bandwidth of $\sim 10^3$ Hz, the number of pulsars in each mHz frequency bin is $\sim 10$ for $f > 100$~Hz. In the low frequency regime this number is much larger. Thus, it is not possible to resolve the signal from each pulsar, the signals from the pulsars create a continuum.

In this work, we are mainly concerned about pulsars which emit in the frequency band of the ground based laser interferometers, which correspond to a time period of milliseconds. Hence we can only consider the distribution of MSPs, which is essentially dominated by $p_1(\log f_r)$ of the bimodal pulsar distribution. Hence, from this point, we only consider the high frequency peak of the pulsar distribution and ignore the low frequency peak altogether. Also, for brevity, we omit the subscript from $\mu_1$, $\sigma_1$ and replace them by $\mu$, $\sigma$ respectively.

Considering a log-normal distribution of sources presented in Eq.~(19) of \citet{hotspot} (and ignoring the low frequency sources), one can then write
\begin{eqnarray}
\label{eq:nfratio}
 &&\frac{N(f(1+z))}{N(f)} \ = \ \frac{1}{1+z} \ \frac{p_1(\log{((1+z)f/2))}}{p_1(\log{(f/2)})} \\
 && \quad = \ \frac{1}{1+z} \exp{\left[-\log\left(\frac{(1+z)f^2}{4\mu^2}\right) \frac{\log(1+z)}{2\sigma^2}\right]} \, .
 \nonumber
\end{eqnarray}
%
%
Then the final expression, eqn.~(\ref{eq:ratIsoVirgo}), becomes
\begin{eqnarray}
\label{eq:genHIso}
&& H_\iso(f) \ = \ H_\Virgo(f) \ \left( \frac{\DelOm_\Virgo \, d_\Virgo^2}{N_\Virgo} \right)
\ \frac{c}{H_0}  \times \\
&& \qquad \int_0^\infty \d z \, n(z) \, \frac{(1+z)^4}{E(z)} \, e^{-\log\left(\frac{(1+z)f^2}{4\mu^2}\right) \frac{\log(1+z)}{2\sigma^2}} \,. \nonumber
\end{eqnarray}

Now, again, to proceed further we will have to introduce cosmology. As before, we use $n(z)$ given by eq.~(\ref{eq:nofz}) for two different cosmological models.
\begin{enumerate}
\item Universe without dark energy

As was in the case of total flux comparison, the justification for including this case is to have analytical handle without deviating too much from the realistic regime. Substituting $E(z) = (1 + z)^{3/2}$ in eqn.~(\ref{eq:genHIso}) one gets
\begin{equation}
\label{eq:noLambdaHIsoInfZ}
\begin{split}
H_\iso(f) \ = & \ H_\Virgo(f) \ \left( \frac{n_0 \, \DelOm_\Virgo \, d_\Virgo^2}{N_\Virgo} \right)
\ \frac{c}{H_0}  \times \\
& \quad\int_0^{z_\text{max}}\d z \, (1+z)^{5/2} \, e^{-\log\left(\frac{(1+z)f^2}{4\mu^2}\right) \frac{\log(1+z)}{2\sigma^2}} \, ,\\
= & \ H_\Virgo(f) \ \left( \frac{n_0 \, \DelOm_\Virgo \, d_\Virgo^2}{N_\Virgo} \right) \ \frac{c}{H_0}  \, (\sigma \ln 10) \, \times \\
 &  \quad e^{y^2/2} \sqrt{\pi \over 2} \left[\erf\left(\frac{y_m - y}{\sqrt{2}}\right) \ + \ \erf\left(\frac{y}{\sqrt{2}}\right)\right] \,,
\end{split}
\end{equation}
where,
\begin{eqnarray}
y &:=& \frac{7}{2}\sigma\ln 10 \ - \ \frac{1}{\sigma\ln 10}\ln\left(\frac{f}{2 \mu}\right) \, ,\\
y_m &:=& \frac{1}{\sigma}\log(1 + z_\text{max}) \, .
\end{eqnarray}
Here we use the standard definition of the error function
\begin{equation}
\erf(x) \ := \ \frac{2}{\sqrt{\pi}}\int_0^{x} \, e^{-t^2} \, \d t \, ,
\end{equation}
which happens to be an odd function:
\begin{equation}
\erf(-x) \ := \ \frac{2}{\sqrt{\pi}}\int_0^{-x} \, e^{-t^2} \, \d t  \ = \ - \erf(x) \, .
\end{equation}

In the limit $z_\text{max} \rightarrow \infty$ (that is, $y_m \rightarrow \infty$), for finite values of $f$, $\erf((y_m - y)/\sqrt{2})  \rightarrow 1$. Hence, in that case, one can write
\begin{eqnarray}
\label{eq:noLambdaHIso}
&& H_\iso(f) \ = \ H_\Virgo(f) \ \left( \frac{n_0 \, \DelOm_\Virgo \, d_\Virgo^2}{N_\Virgo} \right)
\ \frac{c}{H_0}  \times \\
&& \qquad \left[  \sqrt{\pi/2} \, (\ln 10) \, \sigma \, e^{y^2/2} \left(1 \ + \ \erf(y/\sqrt{2})\right) \right] \,. \nonumber
\end{eqnarray}
Note that in this special case, $H_\iso(f)/H_\Virgo(f)$ is a monotonically decreasing function of frequency.

\item $\Lambda$CDM cosmology    

In this case we substitute eqn.~(\ref{eq:defE}) in eqn.~(\ref{eq:genHIso}) and then numerically integrate over redshift $z$ for each frequency bin $f$:
\begin{eqnarray}
\label{eq:LCDMHIso}
&& H_\iso(f) \ = \ H_\Virgo(f) \ \left( \frac{n_0 \, \DelOm_\Virgo \, d_\Virgo^2}{N_\Virgo} \right)
\ \frac{c}{H_0}  \times \\
&& \qquad \int_0^{z_\text{max}} \d z \, \frac{(1+z)^4 \, e^{-\log\left(\frac{(1+z)f^2}{4\mu^2}\right) \frac{\log(1+z)}{2\sigma^2}}}{\sqrt{\Omega_\Lambda + (1+z)^3 \Omega_m)}} \,. \nonumber
\end{eqnarray}
\end{enumerate}

In this paper, we use only one set of $H_\Virgo(f)$ taken from \citet{hotspot} which corresponds to the eccentricity $\epsilon = 10^{-5}$. Since $H_\Virgo(f) \sim \epsilon^2$, both $H_\iso(f)$ and $H_\Virgo(f) $ also scale in the same way, so the PSDs and SNRs for other values $\epsilon$ can be readily obtained. Due to this simple scaling, the main results of this paper, which involve ratios of SNRs, are independent of the choice of $\epsilon$. The particular choice of $\epsilon = 10^{-5}$, though much larger than the current belief, was made only for brevity, to keep the SNRs in the range $0.1$-$1$ for easy comparison.

\begin{figure}[h]
\includegraphics[width=0.5\textwidth]{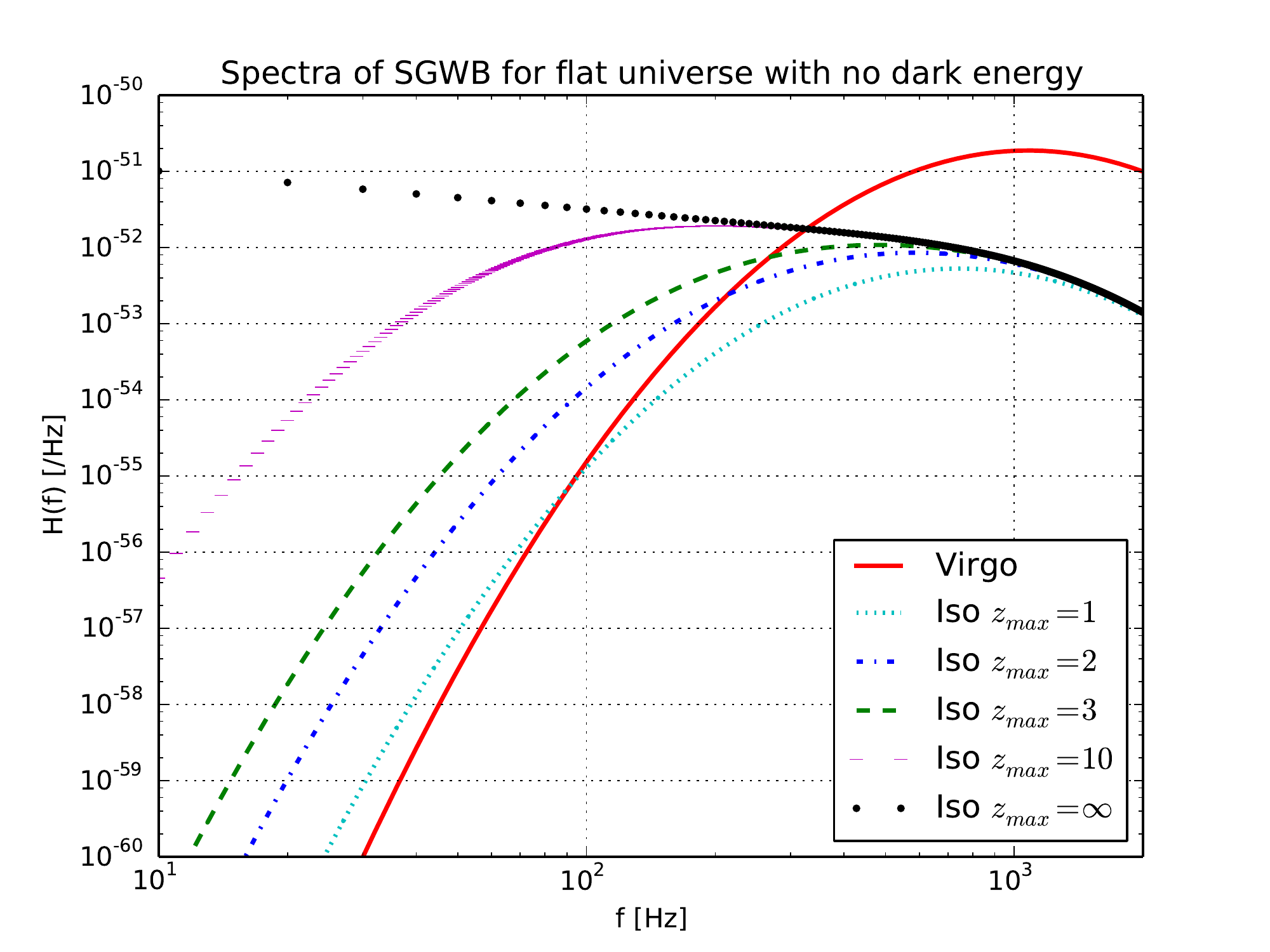}
\includegraphics[width=0.5\textwidth]{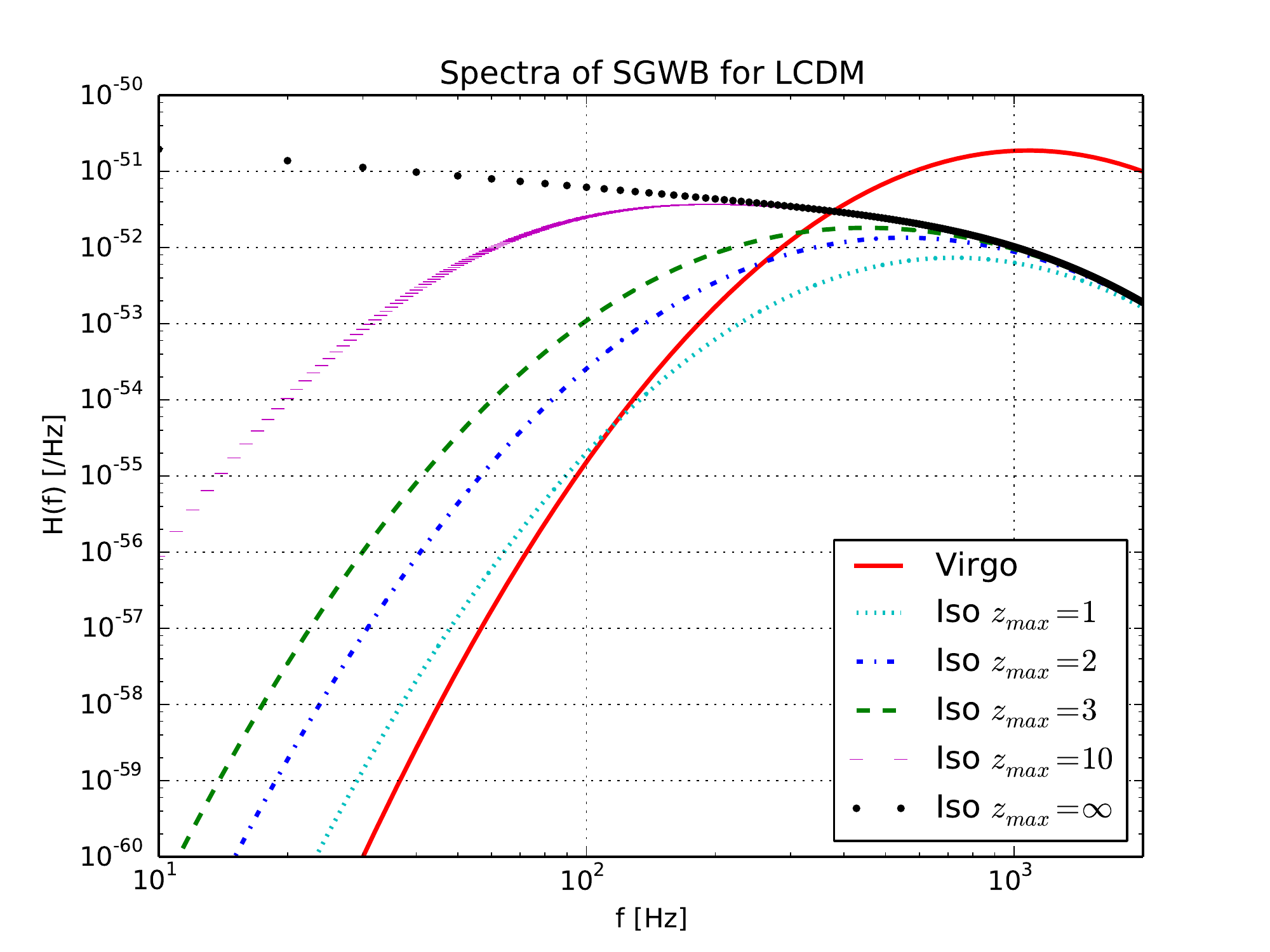}
\caption{The plots show the power spectral densities of SGWB strain from the Virgo cluster, $H_\Virgo(f)$, and the statistically isotropic background, $H_\iso(f)$, {\em in the same solid angle} ($\sim 50$ sq. degrees or $0.012$ steradian) as that of the Virgo cluster for two different cosmologies: no dark energy (above) and $\Lambda$CDM (below) for different choices of $z_{\text{max}}$. The upper and lower panel plots differ by a factor of few, which makes a significant difference in the overall observed SNR.\label{fig:H}}
\end{figure}
Figure~\ref{fig:H} shows the comparison between $H_\Virgo(f)$ and $H_\iso(f)$ for different redshift cut-offs for two cosmological models: no dark energy (above) and $\Lambda$CDM (below). Qualitatively also the results are not surprising. The high redshifts sources move to lower frequencies. Hence, higher the redshift cut-off stronger the PSD at lower frequencies.
Although only small changes can be noticed in the top and bottom panels of Figure~\ref{fig:H}, the differences in the plots are of the order of factor of few, which makes a significant difference in the overall observed SNR.

\section{Observed Signal-to-Noise Ratio}
\label{sec:SNR}

Now we have all the details to answer the central questions---which component of the SGWB is more important in searches for a SGWB, does an optimal search make a significant difference?

\begin{figure*}[t]
\includegraphics[width=0.45\textwidth]{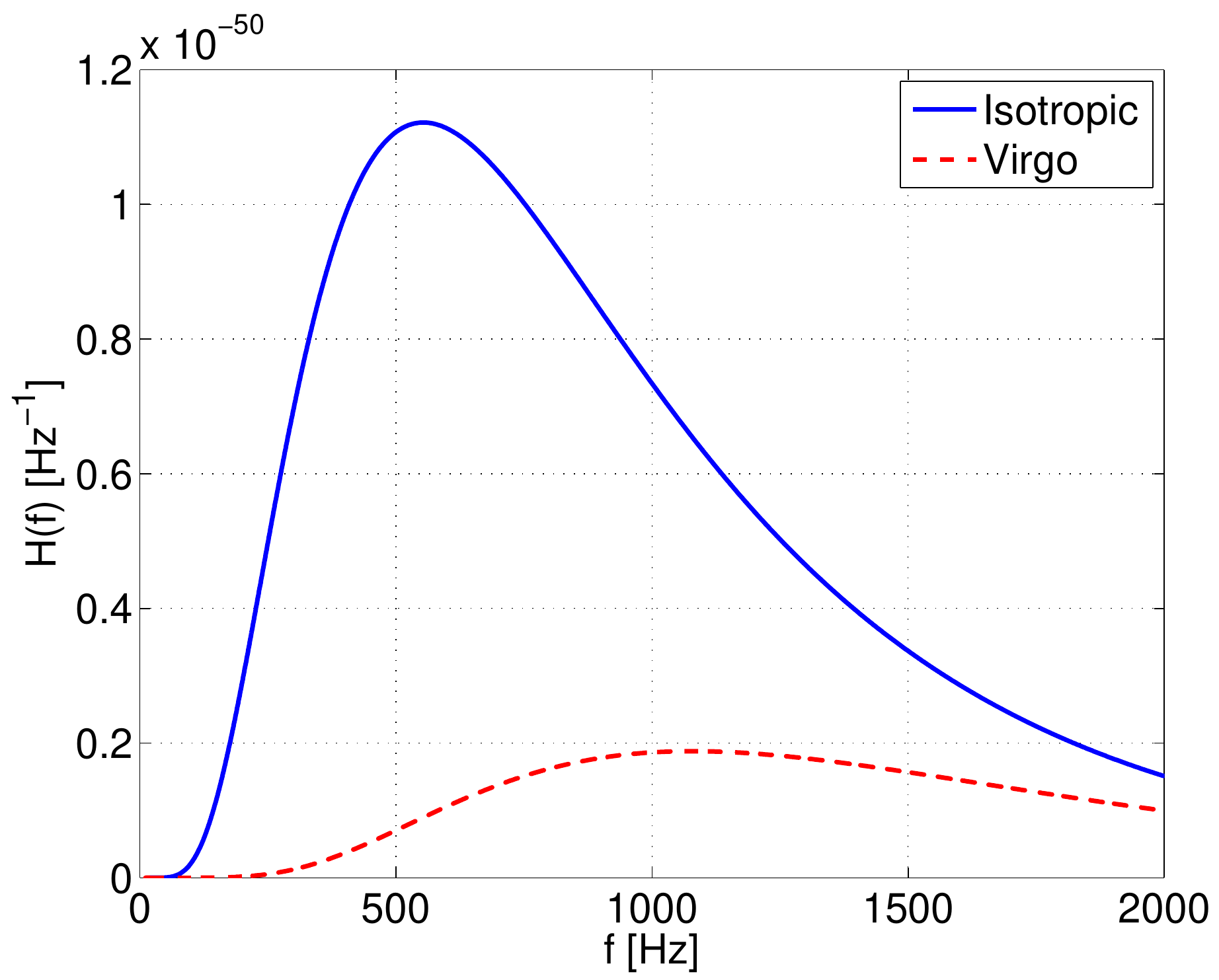}
\includegraphics[width=0.45\textwidth]{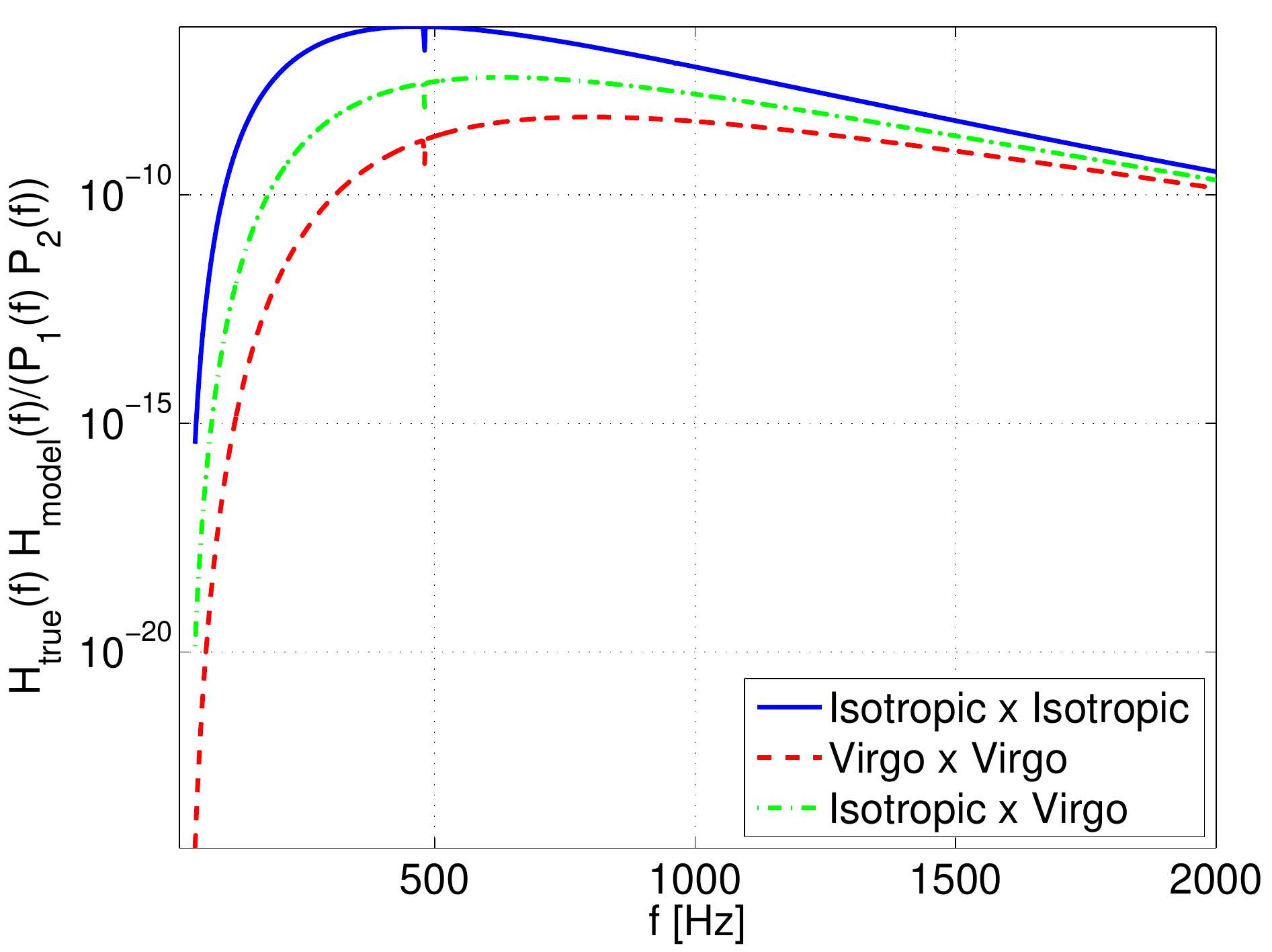}
\includegraphics[width=0.45\textwidth]{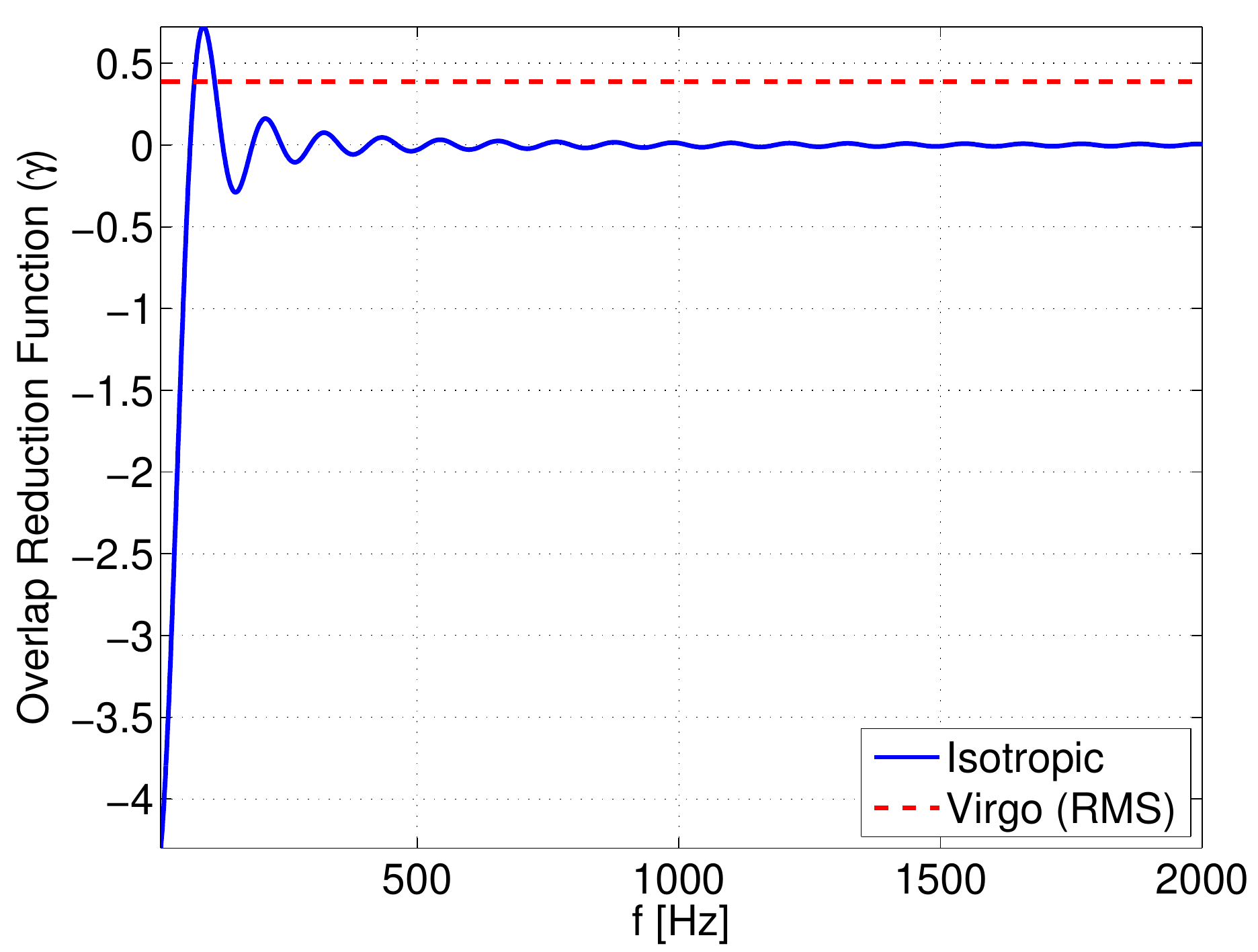}
\includegraphics[width=0.45\textwidth]{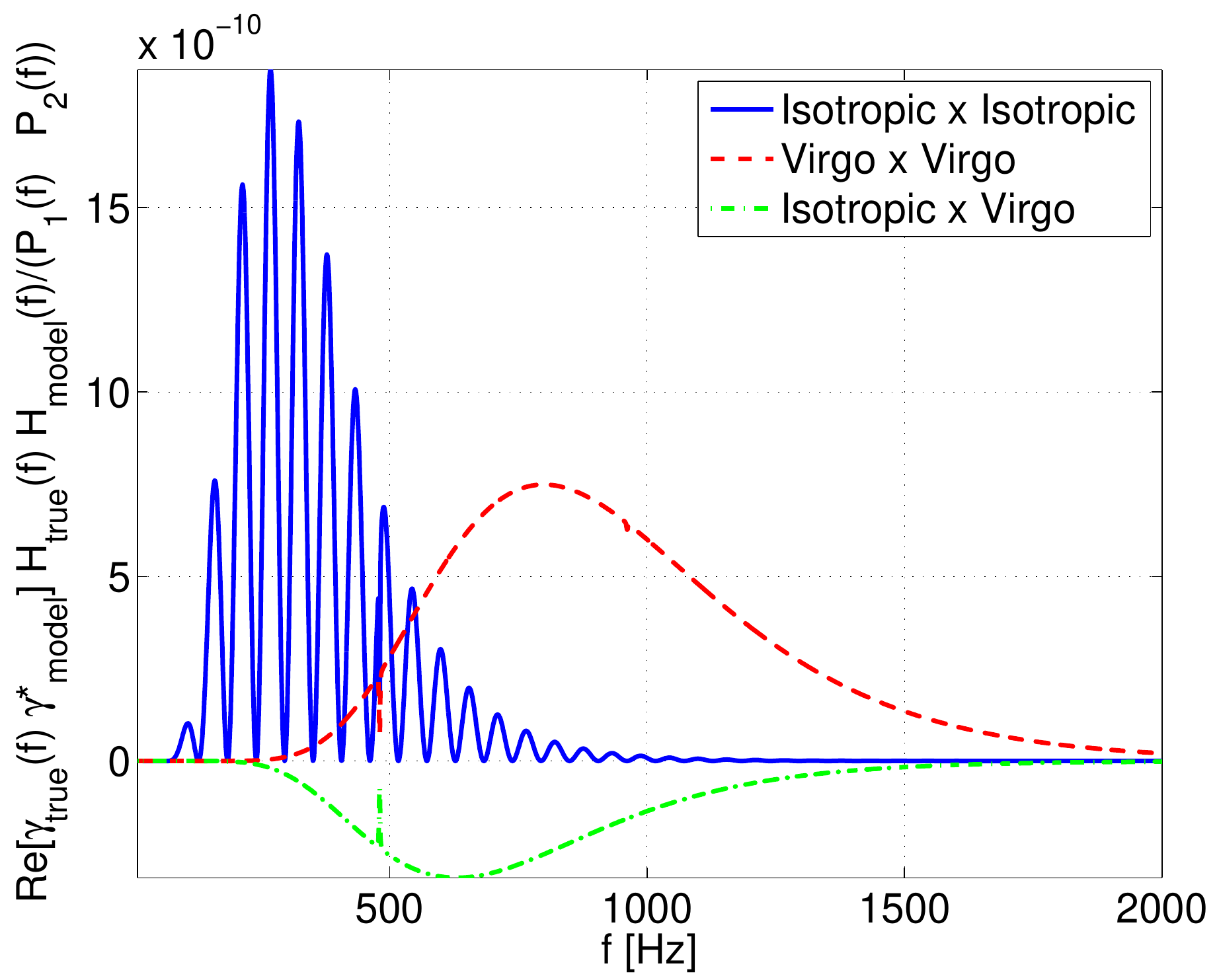}
\caption{This figure shows how different quantities combine to yield small increases in the SNR for directed search for the Virgo cluster (dashed line) and isotropic search (solid line) with $z_\text{max} = 2$. Top left shows $H(f)$ for isotropic background integrated over all sky and Virgo cluster; Top right the frequency spectrum part of the optimal search filter; Bottom left are the overlap reduction function. Bottom right is the final integrand in the expression for SNR, here the square-root of the ratio of the area covered under the solid line to that under the dashed line is the ratio of SNRs (in this case $0.7328$) provided in Table~\ref{tab:SNR}.\label{fig:SNRSplit}}
\end{figure*}

To address these questions we review key results relevant for an anisotropic SGWB. A substantial amount of literature has been written in developing anisotropic searches~\cite{Michelson87, christ92, flan93, allen01, allen97, LIGOT04014000Z, ballmer06, mitra08, mitra09a, Mitra10b}. The key results from these papers can be summarised as follows:
In order to search for an anisotropic background one generally divides the data in many segments with time intervals much greater than the light travel time delay between the detector sites (few tens of milliseconds), but small enough so that the earth can be considered stationary over that period. Generally the period is taken as few tens of seconds~\cite{sgwbS5dir,sgwbS5iso}. The data from pairs of detectors are then correlated in the frequency domain with suitable filters to search for different signals.
It can be shown that, in order to search for an unpolarised SGWB with angular power distribution $\P(\Om)$ and spectrum $H(f)$, the optimal statistic is~\cite{SanjitThesis}
\begin{equation}
S \ = \ \frac{\displaystyle{\sum_{i=1}^n \int_{-\infty}^{\infty} \d f \frac{H(f) \, \gamma_\P^*(t_i,f)}{P_1(t_i;|f|) \, P_2(t_i;|f|)} \, \ft{s}_1^*(t_i;f) \, \ft{s}_2(t_i;f)}}{\displaystyle{\T \sum_{i=1}^n \int_{-\infty}^{\infty} \d f \frac{H^2(f) \, |\gamma_\P(t_i,f)|^2}{P_1(t_i;|f|) \, P_2(t_i;|f|)}}}, \label{eq:GWBS}
\end{equation}
where $t_i$ is the segment time, $\ft{s}_I^*(t_i;f)$ and $P_I(t_i;|f|)$ are respectively the short-term Fourier transform and noise PSD of the data from detector $I$ and,
\begin{eqnarray}
\gamma_\P(t,f) &:=& \sum_{A=+,\times}\int_{S^2} \, \d\Om \, \P(\Om) \times \label{eq:gamma}\\
&& \quad F_1^A(\Om,t) \, F_2^A(\Om,t) \, e^{2\pi \i f \Om\cdot\mathbf{\Delta x}(t)/c}, 
\nonumber
\end{eqnarray}
is the general overlap reduction function expressed in terms of the separation vector between the detectors $\mathbf{\Delta x}(t)$ and the antenna pattern functions $F_I^A(\Om,t)$. The bandwidth of the overlap reduction function for the isotropic search, ${\cal P}(\Om) = 1/4\pi$, is very low, $0$ to $\sim 60$~Hz, for the LIGO Hanford-Livingston (HL) baseline, as shown in figure~\ref{fig:SNRSplit}, which excludes the most sensitive band of the ground based laser interferometric detectors. This unfortunate fact significantly limits the effectiveness in finding astrophysical SGWB generated in the local universe. It may be able to detect high frequency sources if they are appropriately redshifted. The directed search overlap reduction function, on the other hand, has an infinite bandwidth, which means that the search is only limited by the detector bandwidth, and hence can capture a wide range of high frequency sources.

Had the search been optimal, that is, if the model sky $\P(\Om)$ and PSD $H(f)$ matched the true counterparts $\P_\true(\Om)$ and $H_\true(f)$ respectively, the expectation value of the observed signal-to-noise ratio would be~\cite{SanjitThesis}
\begin{equation}
\SNR_\true \ = \ 2 \sqrt{\T \sum_{i=1}^n \int_{-\infty}^{\infty} \d f \frac{H_\true^2(f) \, |\gamma_{\P_\true}(t_i,f)|^2}{P_1(t_i;|f|) \, P_2(t_i;|f|)}} \, .
\end{equation}
For a given background, this is the {\em maximum observable} average SNR. If, however, the model does not match the true sky, (expectation of) the observed SNR becomes
\begin{equation}
\begin{split}
&\SNR_\obs \ =\ 2 \sqrt{\T} \ \times \\
& \quad \frac{\displaystyle{\sum_{i=1}^n \int_{-\infty}^{\infty} \d f \frac{H(f) \, \gamma_\P^*(t_i,f)}{P_1(t_i;|f|) \, P_2(t_i;|f|)} \, H_\true(f) \, \gamma_{\P_\true}(t_i,f)}}{\displaystyle{\sqrt{\sum_{i=1}^n \int_{-\infty}^{\infty} \d f \frac{H^2(f) \, |\gamma_\P(t_i,f)|^2}{P_1(t_i;|f|) \, P_2(t_i;|f|)}}}} \, .
\end{split}
\end{equation}

Using the above formulae we can now perform numerical evaluation for the following quantities to answer the questions we have raised:
\begin{enumerate}
\item $\SNR^\Virgo_\Virgo$:  SNR in a directed search for the Virgo cluster. We put $\P_\true(\Om) = \P(\Om) = \delta(\Om - \Om_\Virgo)$, $H_\true(f) = H(f) = H_\Virgo(f)$ (assuming that the Virgo cluster is nearly a point source for the resolution of the radiometer formed by the LIGO HL baseline).
\item $\SNR^\iso_\iso$: SNR obtained by the isotropic search for the isotropic background. We get this by setting $\P_\true(\Om) = \P(\Om) = 1/4\pi$, $H_\true(f) = H(f) = H_\iso(f)$.
\item $\SNR^\iso_\Virgo$:  SNR obtained by doing a directed search when the actual background is isotropic. We use $\P_\true(\Om) = 1/4\pi$, $\P(\Om) = \delta(\Om - \Om_\Virgo)$, $H_\true(f) = H_\iso(f)$, $H(f) = H_\Virgo(f)$. This number when compared to $\SNR^\Virgo_\Virgo$, will quantify how much the directed search will be able to differentiate between the isotropic and anisotropic search and hence justify if the search is useful for probing the SGWB anisotropy in presence of an isotropic SGWB.
\item $\SNR^\Virgo_\iso$:  SNR contributed by the Virgo cluster to the isotropic search. We put $\P_\true(\Om) = \delta(\Om - \Om_\Virgo)$, $\P(\Om) = 1/4\pi$, $H_\true(f) = H_\Virgo(f)$, $H(f) = H_\iso(f)$. This number shows if the isotropic search will be able to detect the presence of a localised source. By comparing with $\SNR^\Virgo_\Virgo$ one would be able to conclude if a dedicated directed search produces significantly better results.
\end{enumerate}

 Since we are using only one spectrum for the Virgo cluster for this relative study and the spectrum hardly depends on cosmology, we get only one value of $\SNR^\Virgo_\Virgo$ in this paper, which happens to be $\SNR^\Virgo_\Virgo = 0.6067$ for the LIGO HL baseline with an observation time of one sidereal day, advanced LIGO design sensitivity ``of zero-detuning of the signal recycling mirror, with high laser power''~\cite{AdvLIGO-SRD} and frequency range $10 - 2000$~Hz. The values of the other three SNRs, $\SNR^\iso_\iso$, $\SNR^\iso_\Virgo$ and $\SNR^\Virgo_\iso$ are listed in Table~\ref{tab:SNR} for different redshift cut-offs and cosmologies for the same detector pair and observation time.
\begin{table*}[t]
\begin{tabular}{|l|r|r|r|r|r|r|r|r|r|r|r|}\hline\hline
& \multicolumn{5}{|c|}{No dark energy} & \multicolumn{5}{|c|}{$\Lambda$CDM cosmology} \\\cline{2-11}
$z_\text{max} \ \rightarrow$ & $1$ & $2$ & $3$ & $10$ & $\infty$ & $1$ & $2$ & $3$ & $10$ & $\infty$\\\hline\hline
$\SNR^\iso_\iso$ & 0.1049 & 0.2678 & 0.4775 & 2.7007 & 9.2317 & 0.1514 & 0.4446 & 0.8431 & 5.2149 & 17.8521 \\\hline
${\SNR^\iso_\iso}/{\SNR^\Virgo_\Virgo}$ & 0.1729 & 0.4414 & 0.7870 & 4.4515 & 15.2163 & 0.2495 & 0.7328 & 1.3897 & 8.5955 & 29.4249 \\\hline
${\SNR^\iso_\Virgo}/{\SNR^\Virgo_\Virgo}$ & 2.0933e-4 & 3.0328e-4 & 3.3295e-4 & 3.2471e-4 & 3.1647e-4 & 2.8845e-4 & 4.5162e-4 & 5.0766e-4 & 5.1096e-4 & 4.9448e-4\\\hline
${\SNR^\Virgo_\iso}/{\SNR^\Virgo_\Virgo}$ & 1.2148e-3 & 6.8732e-4 & 4.2360e-4 & 7.2853e-5 & 2.0768e-5 & 1.1554e-3 & 6.1645e-4 & 3.6591e-4 & 5.9337e-5 & 1.6812e-5 \\\hline\hline
\end{tabular}
\caption{This table shows the (expected) SNRs and relative SNRs for different combinations of source and model power spectra and sky model of the SGWB observed with LIGO Hanford-Livingston baseline in comparison with directed search SNR for the same baseline for the Virgo cluster is $\SNR^\Virgo_\Virgo = 0.6067$. It can be seen that the directed search for Virgo cluster exceeds the SNR of the all sky integrated search for an isotropic background if the background is dominated by sources below a redshift of $z_\text{max} \sim 3$. The total observation time is taken as one sidereal day, frequency range $10-2000$~Hz and Advanced LIGO design noise PSD of ``zero-detuning of the signal recycling mirror, with high laser power''~\cite{AdvLIGO-SRD}.\label{tab:SNR}}
\end{table*}
The table shows that when the universe has MSPs at high redshifts (younger universe) and it is statistically isotropic at large scales, the isotropic search performs well. However, if the background is dominated by nearby sources (older universe) a localised search can out perform an all sky isotropic search. Moreover, in all the cases the directed search for the isotropic background is much lower, {\em by more than two orders of magnitude} than the directed search for a localised source, signifying that the directed search is highly sensitive to the anisotropy of a background and would be able make a sky map, provided the background is stronger than the detector noise level (integrated over the observation time). Finally, the isotropic search for a localised background is even more suboptimal due to the mismatch in signal and model, the SNR is orders of magnitude below the optimal SNR, hence an isotropic search would not be able to detect the presence of a localised source. Note that the results for the high $z_\text{max}$ are provided for academic interest, as the $n(z) = n_0$ model is bound to fail for high values of $z_\text{max}$.

The reason why the isotropic background, which only dominates in a small range of low frequencies, can still compete with the Virgo cluster, is illustrated in figure~\ref{fig:SNRSplit} obtained for the $\Lambda$CDM cosmology with $z_\text{max} = 2$. The top left panel shows $H(f)$ for isotropic and Virgo backgrounds; the top right panel shows $H_\true H_\obs /(P_1(f) P_2(f))$; the bottom left panel shows the overlap reduction function for the two cases; the bottom right plot shows essentially the final integrand which provides the SNR. Even though $H_\iso(f)$ seems to be dominating over $H_\Virgo(f)$ in this case, the large bandwidth of the directed search equalises the observed SNR in either cases when the model matches the true signal. As shown in Table~\ref{tab:SNR}, $\SNR^\Virgo_\Virgo \approx \SNR^\iso_\iso$ for $z_\text{max} = 2$.

For the purpose of completeness, we also include numerical results for the Hanford-Virgo and Livingston-Virgo detector baselines. We do this for a few realistic values of $z_\text{max}$ for $\Lambda$CDM cosmology and Advanced Virgo design sensitivity~\cite{AdvVirgo-SRD}. The results are shown in Table~\ref{tab:LIGOVirgoSNR}. Clearly, for the LIGO-Virgo baselines the results are similar to the LIGO HL baseline. However, isotropic search with the LIGO-Virgo baselines has better high frequency sensitivity~\cite{VirgoORF} than the LIGO HL baseline. This provides comparatively higher isotropic search SNR even for low redshift cut-offs ($z_\text{max} \lesssim 1$).
\begin{table*}[t]
\begin{tabular}{|l|r|r|r|r|r|r|r|r|r|r|r|r|}\hline\hline
& \multicolumn{6}{|c|}{LHO-Virgo} & \multicolumn{6}{|c|}{LLO-Virgo} \\\cline{2-13}
$z_\text{max} \ \rightarrow$ & $0.5$ & $1$ & $2$ & $3$ & $10$ & $\infty$ & $0.5$ & $1$ & $2$ & $3$ & $10$ & $\infty$\\\hline\hline
$\SNR^\Virgo_\Virgo$ & \multicolumn{6}{|c|}{0.1013} & \multicolumn{6}{|c|}{0.1359} \\\hline
$\SNR^\iso_\iso$ & 0.0428 & 0.1295 & 0.3973 & 0.7135 & 2.5182 & 5.3624 & 0.0523 & 0.1581 & 0.4848 & 0.8704 & 3.0615 &  6.4597 \\\hline
$\SNR^\iso_\Virgo$ & 1.47e-5 & 2.23e-5 & -4.35e-5 & -1.81e-4 & -6.82e-4 & -7.24e-4 & 5.99e-6 & 2.18e-5 & 7.52e-5 &  1.31e-4 & 2.39e-4 & 2.36e-4 \\\hline
$\SNR^\Virgo_\iso$ & 3.49e-5 & 1.75e-5 & -1.11e-5 & -2.57e-5 & -2.74e-5 & -1.37e-5 & 1.56e-5 & 1.87e-5 & 2.11e-5 & 2.04e-5 & 1.06e-5 & 4.97e-6\\\hline\hline
\end{tabular}
\caption{The (expected) SNR for different combinations of source and model power spectra and sky model of the SGWB observed with baselines formed by one of the LIGO $4$km detectors [LIGO Hanford Observatory (LHO) and LIGO Livingston Observatory (LLO)] and the Virgo detector whose armlength is $3$km. Here we use $\Lambda$CDM cosmology, total observation time of one sidereal day, frequency range $10-2000$~Hz and design noise power spectrum for the advanced detectors. This table shows that for these baselines directed search for Virgo cluster exceeds the SNR of the all sky integrated search for an isotropic background if the background is dominated by nearby sources with redshift $z_\text{max} \lesssim 1$.\label{tab:LIGOVirgoSNR}}
\end{table*}

\section{Conclusion}
\label{sec:conclusion}

Over the past two decades the search algorithms for both isotropic and anisotropic stochastic gravitational wave background have been developed and applied to data from the ground based Laser Interferometric  GW detectors. In this work we showed that both of these searches are important for different backgrounds and one search cannot replace the other. However, for detection purposes, one can conclude that the isotropic search is more effective when the sources are distributed uniformly across the sky and up to a high redshift, while the directed search is more efficient when most of the sources are in the nearby universe and their distribution is anisotropic, with one (or few) localised sources.

We show that the directed search is highly sensitive to the anisotropy of a background which is useful for making a sky map of the background. The performance of the isotropic search, on the other hand, depends on the expansion history of the universe and population distribution of MSP.  The competition is such that the SNR of the directed search for the Virgo cluster can become comparable or even exceed the all sky integrated isotropic search depending on the cosmology. We have also presented a detailed analysis to show how this competition becomes close even though the all sky integrated background is the result of sources whose number is many orders of magnitude larger. The contrast between different searches can become much stronger for searches over narrower frequency bands.

In this paper, we studied the simplest form of an anisotropic background, namely a highly localised source, compared to an isotropic background. This was sufficient to provide motivation for the directed search. However, one could also perform an explicit comparison like this for other kinds of anisotropic searches, e.g., the spherical harmonic search~\cite{mitra09a}. Different sources with very different population distribution over redshift [$n(z)$] and frequency spectra [$L J(f)$] could also lead to significantly different results. Hence, it may be worth repeating our calculations for different cosmological (no dark energy vs. $\Lambda$CDM), population and spectral models for the sources. These studies would be useful if one is planning to perform specific searches for a certain kind of anisotropy or constituent sources. Note that the general SNR formulae presented in this paper can incorporate any $n(z)$ and $J(f)$, hence complicated population synthesis models can also be included in this formalism.

 The large values of isotropic search SNRs estimated for high redshift cut-offs ($z_\text{max}$) in Tables \ref{tab:SNR} \& \ref{tab:LIGOVirgoSNR} are reflections of the fact that we used a top-hat model for $n(z)$, a model that can not be stretched to high $z_\text{max}$ values. Since these SNRs are highly sensitive to various source and cosmological models, one can even consider studying if the searches would be able to constrain these models (e.g., put an upper limit on $z_\text{max}$ assuming a certain $n(z)$ model).
 
The analyses in this paper are restricted to either highly localised or isotropic backgrounds, which are, by their intrinsic nature, suboptimal. They are currently performed this way due to the lack of reliable models of the anisotropic sky. It may however be possible to gather information from electromagnetic astronomy and construct more optimal filters to perform search for anisotropic backgrounds. Here one may even consider using different spectra for different directions, as the composition of sources in different direction may be different. However, such an analysis does not exist in literature and, also, it may not be very straightforward to develop such an analysis in a computationally viable way. In such cases also, a comparative study similar to this paper needs to be performed, to decide if such an effort is worthwhile.

In this paper we computed most of the numerical results for the LIGO Hanford-Livingston (HL) baseline, which is sufficient to justify the importance of different searches. For academic interest, we also included results for longer baselines including the Virgo detector. However, for longer baselines the resolution of the radiometer is considerably high and the assumption that the Virgo cluster is a point source breaks down. So to get more rigorous SNR estimations, one should consider the finite size of the Virgo cluster and, perhaps, use a model for the mass distribution.

The goal of this paper is not to address the issue of template mismatch and suggest a solution. However, since some of the SNRs in the tables are negative, we must caution the readers that template mismatch can cause negative values of the statistic in specific frequency bands and eventually lead to wrong upper limits when integrated over all frequencies. The absolute value of the estimator may be one way to address the issue, though care must be taken to ensure that the statistic is indeed estimating the desired quantity. This issue can be ignored when one is trying to probe a background that is by far the strongest in the sky in the frequency band considered. In this particular situation the filter is near optimal. Which may not be the case in general. A more rigorous strategy must be developed in order to alleviate this problem, perhaps by fitting all the SGWB components together.

 \section*{Acknowledgements}
 
 We would like to thank Bruce Allen, Stefan Ballmer, Dipankar Bhattacharya, Charles Jose, Dipanjan Mukherjee, Somak Raychaudhury, Joe Romano, Dipongkar Talukder, Eric Thrane, John Whelan and the stochastic group of the LIGO-Virgo Scientific Collaboration for useful discussions. NM is supported by the DST-MPG Max Planck Partner Group Grant (funded by the Department of Science and Technology, India and Max Planck Society, Germany) under the Grant no IGSTC/MPG/PG (AP) 2011. SM acknowledges the support of Department of Science and Technology (DST), India for the SERB FastTrack grant SR/FTP/PS-030/2012. SVD would like to thank IUCAA for a Visiting Professorship.
 
 \bibliography{isoVsLocal}
 
\end{document}